\documentclass[fleqn,10pt]{wlscirep}
\usepackage[utf8]{inputenc}
\usepackage[T1]{fontenc}
\title{Automata representation of successful strategies for social dilemmas}

\author[1]{Yohsuke Murase}
\author[2,*]{Seung Ki Baek}
\affil[1]{RIKEN Center for Computational Science, Kobe, Hyogo 650-0047, Japan}
\affil[2]{Department of Physics, Pukyong National University, Busan 48513, Korea}

\affil[*]{seungki@pknu.ac.kr}

\begin{abstract}
In a social dilemma, cooperation is collectively optimal, yet individually each
group member prefers to defect. A class of successful strategies of direct
reciprocity were recently found for the iterated prisoner's dilemma and for the
iterated three-person public-goods game: By a successful strategy, we mean that
it constitutes a cooperative Nash equilibrium under implementation error, with
assuring that the long-term payoff never becomes less than the co-players'
regardless of their strategies, when the error rate is small. Although we have a
list of actions prescribed by each successful strategy, the rationale behind
them has not been fully understood for the iterated public-goods game because
the list has hundreds of entries to deal with every relevant history of previous
interactions. In this paper, we propose a method to convert such history-based
representation into an automaton with a minimal number of states. Our main
finding is that a successful strategy for the iterated three-person public-goods
game can be represented as a $10$-state automaton by this method. In this
automaton, each state can be interpreted as the player's internal judgement of
the situation, such as trustworthiness of the co-players and the need to redeem
oneself after defection. This result thus suggests a comprehensible way to
choose an appropriate action at each step towards cooperation based on a
situational judgement, which is mapped from the history of interactions.
\end{abstract}
\begin{document}

\flushbottom
\maketitle
%
%
\thispagestyle{empty}

\section*{Introduction}
George Berkeley says that a man who believes in no future state has no
reason to postpone his own private interest or pleasure to doing his
duty~\cite{chalmers2008british}.
Reciprocity is one way to establish
cooperation between rational individuals under this shadow of
future~\cite{nowak2006evolutionary,sigmund2010calculus,van2012direct,hilbe2018partners}.
Tit-for-tat (TFT) is one of the most popular reciprocal strategies in the
iterated prisoner's dilemma (PD) game~\cite{axelrod1984evolution}. Just by
replicating the co-player's previous action, it embodies several, intuitively
appealing properties, that is, being clear, nice, provokable, and forgiving.
However, if the prescribed actions are misimplemented, two TFT players easily run
into TFT retaliation~\cite{molander1985optimal,boyd1989mistakes}, so the
long-run average payoff becomes as low as those between two RANDOM players,
where a RANDOM strategy means choosing cooperation with probability $1/2$.
Moreover, a TFT population is invaded by unconditional cooperators
because a TFT player cannot distinguish an unconditional cooperator from another
TFT player.
Generous TFT has been suggested to avoid the TFT
retaliation~\cite{nowak1992tit,imhof2005evolutionary,imhof2007tit,imhof2009stochastic},
but it is outperformed by Win-Stay-Lose-Shift
(WSLS)~\cite{kraines1989pavlov,nowak1993strategy}.
WSLS also solves the problem of distinguishability in the sense that it earns a
strictly higher average payoff against an unconditional cooperator.
However, it is vulnerable against unconditional defectors.

A notable progress in the iterated PD game is the discovery of the
zero-determinant (ZD) strategies~\cite{press2012iterated}. Each of them is a
memory-one strategy, generally stochastic, and it can enforce a certain linear
relationship between its own payoff and co-players' payoffs irrespective of the
co-player's strategy~\cite{ichinose2018zero,mamiya2019strategies}. This is true
even when the co-player has a longer memory or when the strategy is known to the
others. When both the players attempt to extort each other using an extortionate
ZD strategy, they end up with mutual defection, so an extortionate strategy is
hard to evolve as a
group~\cite{hilbe2013evolution,stewart2013extortion,hilbe2013adaptive,szolnoki2014evolution}.
TFT is a special case of the ZD strategies, equalizing the players' payoffs in
the long run.
The ZD strategies have been studied not only in a well-mixed population but also
in structured ones~\cite{szolnoki2009phase,szolnoki2014defection} because of the
importance of spatiotemporal dynamics from a statistical-physical
viewpoint~\cite{perc2017statistical}.

To explore an even stronger class of strategies, researchers have proposed a
strategy called TFT-anti-TFT (TFT-ATFT), which can be understood as a
modification of TFT~\cite{yi2017combination}. It has been devised to remedy the
problems of TFT by satisfying the following three criteria:
\begin{enumerate}
\item Efficiency: If all the players in the game have adopted this strategy in
common, they will reach mutual cooperation with probability one as the
implementation error rate $e$ approaches zero.
\item Defensibility: If the focal player uses this strategy, her expected payoff
is greater than or equal to any of her co-players' regardless of their
strategies.
\item Distinguishability: If all the co-players are unconditional cooperators,
the expected payoff from this strategy is strictly higher than theirs.
\end{enumerate}
Here, an implementation error (also called execution error or mistake) refers to an event that a player erroneously takes the opposite action to the prescription of the strategy.
Unlike the perception error, it is assumed that all the players, including the one who committed an error, correctly perceive which actions are actually taken.
The class of strategies satisfying these three criteria are called successful
hereafter. The first two criteria are especially important because a cooperative
Nash equilibrium is formed when efficiency and defensibility criteria are
simultaneously satisfied~\cite{yi2017combination}. Moreover, it is guaranteed
that the focal player's long-term payoff will never be less than those of the
others against any kind of strategies. Just as is the
case for the ZD strategies, this relation is assured even when the co-players
have a longer memory length or when they know the focal player's deterministic
strategy. TFT-ATFT is a memory-two strategy, namely, it prescribes its next
action depending on the history profile for previous two rounds.
The definition of TFT-ATFT is given in Table~\ref{tab:tft-atft}.
As indicated by the name, it is a combination of TFT and ATFT:
If it correctly played TFT for the two previous steps, it keeps playing TFT.
Otherwise, it behaves as ATFT. If mutual cooperation is reached,
or if the co-player unilaterally defects twice in a row, it is time to go back to TFT.
Thus, when the player erroneously deviates from TFT, the ATFT part is activated for a while
to correct the error, whereby mutual cooperation can be made robust in a noisy environment
without violating defensibility.
Regarding efficiency, we mention that perception error can also be corrected if
it occurs with a much longer time scale than implementation errors~\cite{yi2017combination}.

\begin{table}[htbp]
\caption{
List of actions in TFT-ATFT~\cite{yi2017combination}. Two players Alice and Bob are involved in the iterated PD game, and we assume that Alice is playing TFT-ATFT. Let $A_t$ and $B_t$ denote Alice's and Bob's actions at time $t$, respectively. In this table, each state means $(A_{t-2} A_{t-1}, B_{t-2} B_{t-1})$, and the corresponding action means $A_t$.
}\label{tab:tft-atft}
\begin{center}
\begin{tabular}{cccc}
\hline
State & Action & State & Action \\
\hline
$(cc,cc)$ & $c$ & $(dc,cc)$ & $c$ \\
$(cc,cd)$ & $d$ & $(dc,cd)$ & $d$ \\
$(cc,dc)$ & $c$ & $(dc,dc)$ & $c$ \\
$(cc,dd)$ & $d$ & $(dc,dd)$ & $c$ \\
$(cd,cc)$ & $d$ & $(dd,cc)$ & $d$ \\
$(cd,cd)$ & $c$ & $(dd,cd)$ & $c$ \\
$(cd,dc)$ & $c$ & $(dd,dc)$ & $c$ \\
$(cd,dd)$ & $d$ & $(dd,dd)$ & $d$ \\
\hline
\end{tabular}
\end{center}
\end{table}

Successful strategies exist not only for the iterated PD game but also for an
iterated public-goods (PG) game~\cite{boyd1988evolution}. The payoff matrix of
the three-person PG game is given as follows:
\begin{equation}
M \equiv \left(
  \begin{array}{c|ccc}
    \ & 0 & 1 & 2  \\\hline
    C & \rho & \frac{2}{3}\rho & \frac{1}{3}\rho \\
    D & 1+\frac{2}{3}\rho & 1+\frac{1}{3}\rho & 1
  \end{array}
\right),
\end{equation}
where the number of defectors among the two co-players is written at the top of
each column , and $\rho$ is a multiplication factor satisfying $1< \rho < 3$.
This is a generalization of the iterated PD game to a three-person case. As in
the PD game, the only Nash equilibrium of the one-shot PG game is full defection
with payoff $M_{D,2} = 1$, which is the worst for the society as a whole. For
the iterated three-person PG game, it has been found that at least
$256$ successful strategies exist in the memory-three strategy
space~\cite{murase2018seven} and that no such strategy exists if the memory
length is less than three.
This fact immediately poses a problem on its understandability: Recall that a memory-three strategy is defined by an action table having $512$ entries
because the number of possible history profiles is $2^{3\times 3} = 512$.

The purpose of this paper is to interpret the successful strategies by
representing them as automata. In the previous works, we have
represented successful strategies in a `history-based' manner so that the next
action is given as a function of the history profile for the last $m$ rounds.
However, a strategy may also be defined as an
automaton~\cite{rubinstein1986finite}, i.e., in a way that
a player has a finite number of
internal states. A player's internal state determines her next action,
and it changes according to the actions taken by the players of the game.
We will show that the decision mechanism behind the actions prescribed by the
strategy can be understood more clearly in this `state-based' representation.

The paper is organized as follows: In the next section, we present an algorithm
to convert a history-based representation to a state-based one. Then, its
applications to some successful strategies will be demonstrated.
We discuss possible interpretations for the resulting
internal states and summarize this work in the last section.

\section*{Method}

\subsection*{Algorithm}
In this section, we show how a history-based strategy can be converted to a
state-based representation. In general, history-based strategies may be regarded
as a subset of state-based ones because one may also regard the history profile
over the previous $m$ rounds as an internal state. In this naive
reinterpretation, the number of states (i.e., history profiles) would amount to
$2^{nm}$, where $n$ is the number of players. Let us consider a directed graph
with $2^{nm}$ nodes, in which each node denotes a distinct history profile and
each link means a transition between a pair of states. Each node has $2^{n-1}$
outgoing links, corresponding to the possible number of actions taken by the
$n-1$ co-players, because the focal player's action has already been fixed by
the strategy under consideration. Note that this graph does not include
transitions caused by implementation error.

\begin{figure}
\begin{center}
\includegraphics[width=0.9\textwidth]{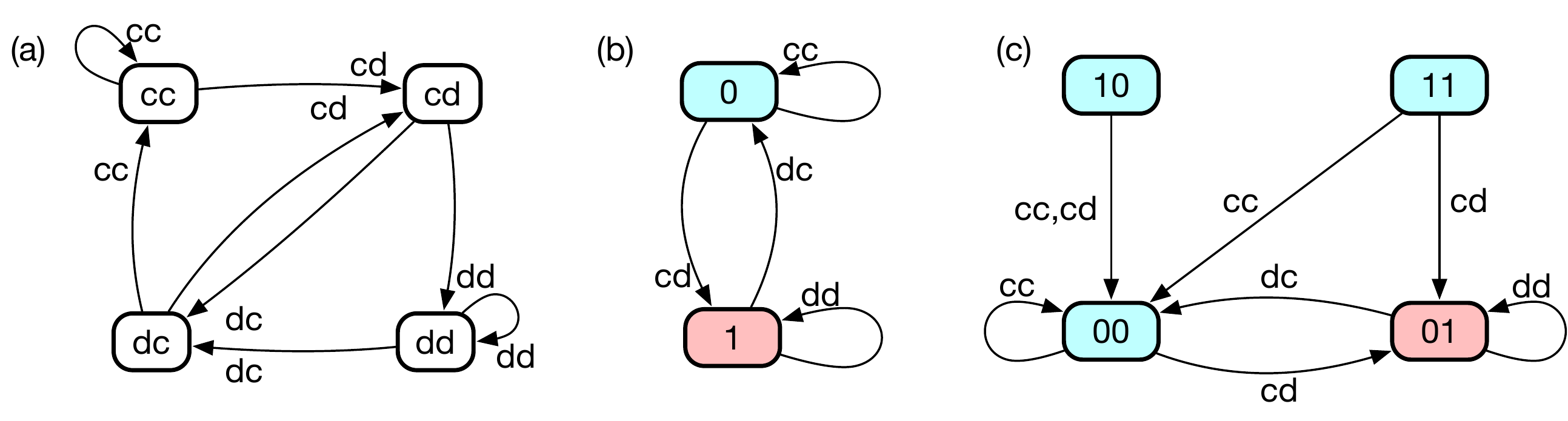}
\end{center}
\caption{
(a) Transition among history profiles of TFT.
Each node is labelled by a history profile, which is a $2$-tuple composed of the
last actions of the two players in this memory-one strategy.
A history profile may also be regarded as an internal state of the focal player
in this naive representation.
Each node has two outgoing links because it has two possible destinations
depending on the co-player's choice between $c$ and $d$.
(b) State-based representation of TFT with two internal states. If the
co-player cooperates (defects), the internal state becomes 0(1), and the focal
player chooses an action based on this state.
The colour of each node indicates the action prescribed at each state: Blue and
red mean cooperation and defection, respectively.
(c) Graph representation of CTFT, one of the most well-known strategies
based on internal states called standing.
Each player's standing is either good (0) or bad (1) from the focal player's
viewpoint. For example, `01' means that the focal player assigns good standing
to herself and bad standing to her co-player.
}
\label{fig:tft_as}
\end{figure}

An example of such a graph is shown in Fig.~\ref{fig:tft_as}(a). Because TFT is
a memory-one strategy for a two-players game, it has four nodes, labelled by
$cc$, $cd$, $dc$, and $dd$, respectively.
Suppose that the current history profile is $cc$. For example, if the
two players' last actions are $c$ and $d$, respectively, the next history
profile becomes $cd$. In case of TFT, the action tuples such as $cd$, denoted as
link attributes, happen to look similar to node labels, but this is not the
case for general memory-$m$ strategies if $m > 1$. Although this representation
fully defines the strategy, it has redundancy. For instance, it is obvious that TFT
can also be represented by a graph with two states as shown in
Fig.~\ref{fig:tft_as}(b)~\cite{van2012direct,hilbe2018partners}.
In case of TFT, it is straightforward to construct the graph in
Fig.~\ref{fig:tft_as}(b) based on Fig.~\ref{fig:tft_as}(a). However, it suddenly
becomes complicated when the memory length gets longer because the number of
nodes grows exponentially.

\begin{figure}
\begin{center}
\includegraphics[width=0.9\textwidth]{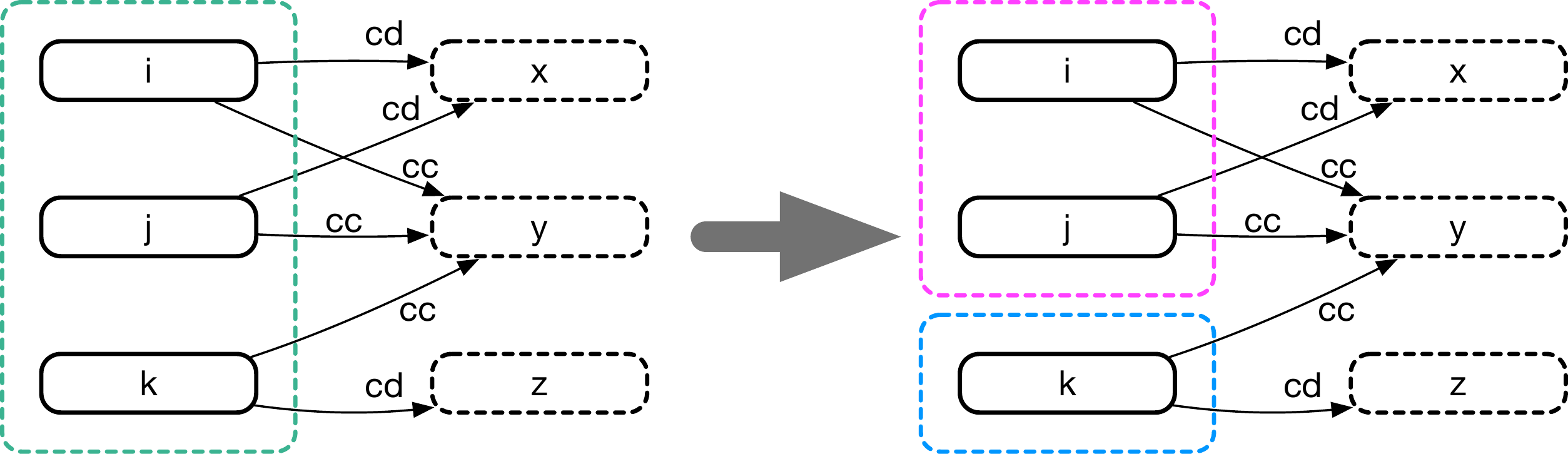}
\end{center}
\caption{
DFA minimization. Nodes $i$, $j$, and $k$ should be split into
two sets, $\{i,j\}$ and $\{k\}$, because $i$ and $j$ lead to the same
future via either $cc$ or $cd$, whereas $k$ responds differently.
}
\label{fig:merge_nodes}
\end{figure}

\begin{figure}
\begin{center}
\includegraphics[width=0.9\textwidth]{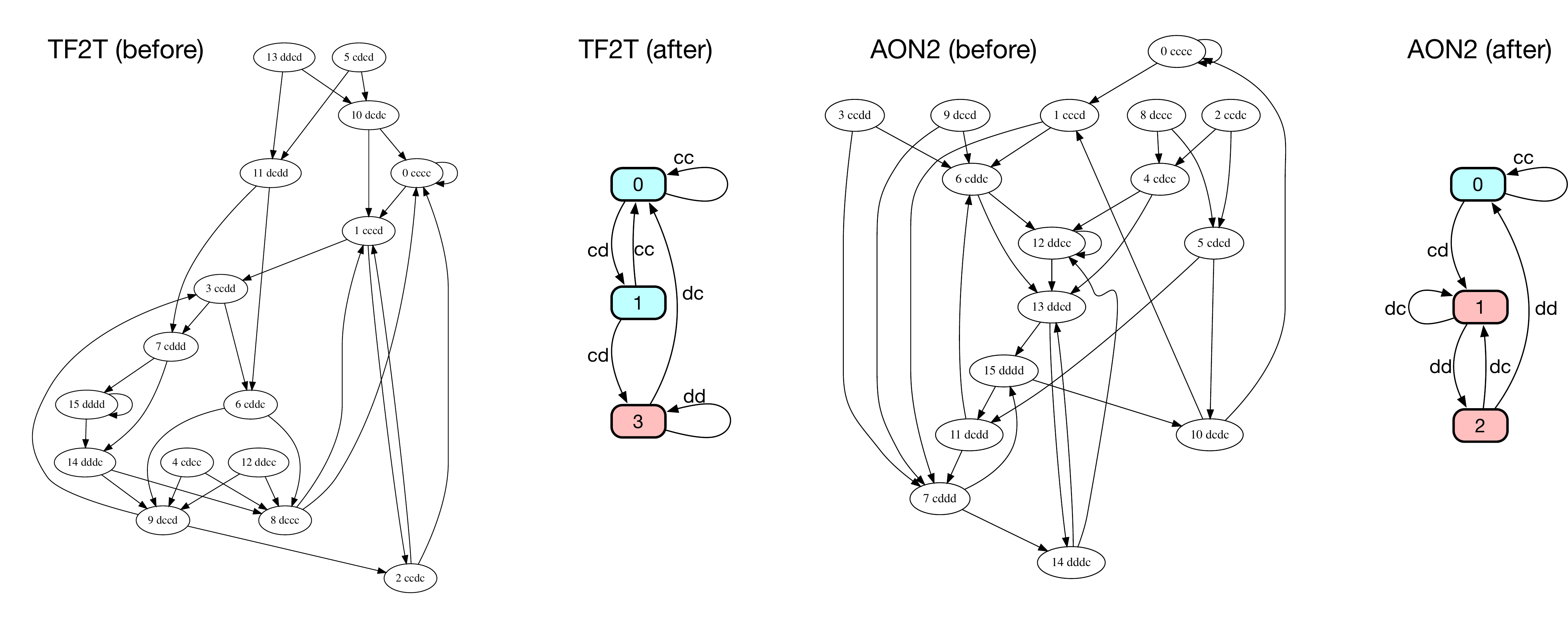}
\end{center}
\caption{
Conversion of history-based representation to state-based one. An example
is Tit-for-Two-Tats (TF2T) before and after the DFA minimization
(left)~\cite{hilbe2018partners}. The
second example on the right hand is AON$_2$, the `all-or-none' strategy for the
PD game among memory-two players~\cite{hilbe2017memory}.
Each of these strategies, which generally has $16$ nodes as a memory-two
strategy, is reduced to an automaton with three internal states by the DFA
minimization. As in Fig.~\ref{fig:tft_as}, blue (red) means that the player
should cooperate (defect) at the state. We have suppressed the action tuples
assigned to the links in the history-based representation for better visibility.
}
\label{fig:conversion_examples}
\end{figure}

Thus, the question is how to simplify a naive representation systematically by
minimizing the number of states. This is known as deterministic-finite-automaton
(DFA) minimization in automata theory~\cite{moore1956gedanken}. Specifically,
we use the following algorithm:
\begin{enumerate}
    \item Define a partition $P_0$ by splitting given states into two sets
    according to their prescription between $c$ and $d$.
    \item Initialize $k = 0$.
    \item Increase $k$ by one.
    For each set in $P_{k-1}$, if a pair of nodes $i$ and $j$ in it are not
    equivalent, divide the set into finer subsets to define a new partition
    $P_k$. Here, nodes $i$ and $j$ are equivalent if the outgoing links from
    these nodes go together in $P_{k-1}$ for any input.
    In our context, an input means an action tuple of the co-players.
    (see Fig.~\ref{fig:merge_nodes}.)
    \item Repeat step 3 until $P_k$ becomes identical to $P_{k-1}$.
\end{enumerate}
In short, we regard two states as identical when they lead to the same future.
The algorithm always terminates after a finite number of steps, and
the final result is uniquely determined irrespective of the order of choosing
node pairs. If we apply this algorithm to Fig.~\ref{fig:tft_as}(a) for instance,
it ends up with two super-nodes $\{cc,dc\}$ and $\{cd,dd\}$, yielding the graph
shown in Fig.~\ref{fig:tft_as}(b) as expected.
The opposite conversion is not always possible. For example, one needs an
infinitely long memory to describe the behaviour of Contrite TFT
(CTFT)~\cite{sugden1986economics} in the history-based
representation~\cite{boerlijst1997logic}, whereas its state-based
version needs only four states [Fig.~\ref{fig:tft_as}(c)].
Figure~\ref{fig:conversion_examples} shows some other examples of the DFA
minimization. It greatly simplifies the graphs, especially when the memory
length is long.

Here, we note that the transitions in Fig.~\ref{fig:conversion_examples} do not take into account the transitions caused by errors.
In other words, the minimized automaton generally loses some of information about erroneous actions while it reproduces the deterministic actions prescribed by the strategy.
In order to fully keep the information of the original history-based representation, one needs to start from the transition graph that has outgoing links corresponding to erroneous actions as well.
An example of such automaton representation will be shown in Fig.~\ref{fig:tft_atft}(c).
In general, we should choose one of the representations depending on our purpose.
While information loss is caused by ignoring error,
the converted automaton representation usually has a smaller number of states, which is helpful in interpreting the strategy.
On the other hand, the full automaton representation keeps all the information of the strategy, making it possible to reconstruct the history-based representation.
In this paper, we mainly take the former approach because our main objective is to better interpret the strategies.

\subsection*{Ethics statement}
No human experiments were conducted in this study.

\section*{Result}

\subsection*{Iterated PD game}

\begin{figure}
\begin{center}
\includegraphics[width=0.9\textwidth]{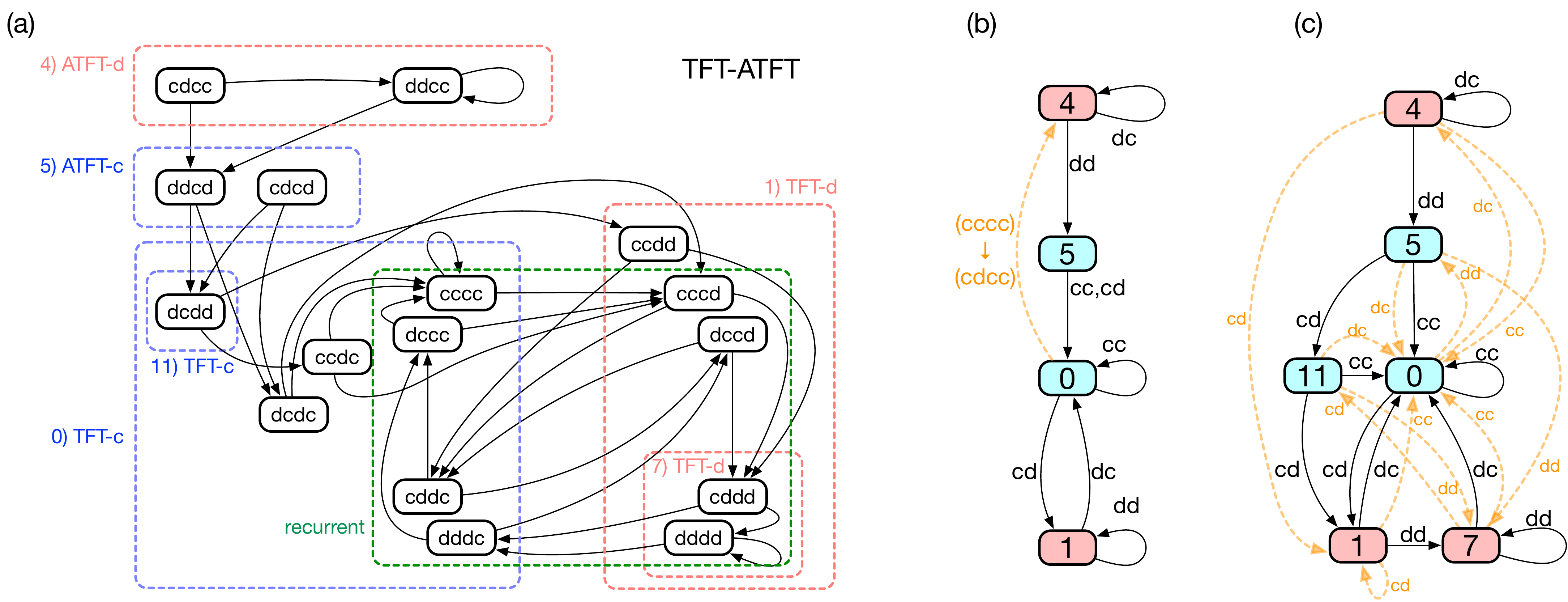}
\end{center}
\caption{
(a) History-based representation of TFT-ATFT. Each node represents a history
profile of the two previous rounds, thus the graph has $16$ nodes in total.
Each node has two outgoing links because the history profile changes depending
on Bob's choice between $c$ and $d$.
The green dashed rectangle shows the strongly connected component responsible
for the TFT behaviour.
(b) State-based representation of TFT-ATFT in which the number of states has
been reduced to four by the DFA minimization algorithm.
As in Fig.~\ref{fig:tft_as}, when the current state is represented by a blue
(red) node, the next action should be cooperation (defection).
The state changes according to the $2$-tuple of actions attached to each link.
For instance, the state changes from `0' to `1' via $cd$, meaning that Alice and
Bob chose $c$ and $d$, respectively, at the previous round.
The four dashed rectangles in red and blue in (a) correspond to the four
nodes in (b). For instance, state `4' in (b) is a super-node formed by merging
$cdcc$ and $ddcc$ in (a). They are $4$ and $12$ in binary, and we have chosen
the former one to denote the super-node.
Likewise, the label of each super-node in (b) originates from the minimum index of its constituent nodes in (a).
This representation is a simplification of (a), thus error-induced transitions are not taken into account.
To show how it handles error occurring with probability of $O(e)$, we denote one of the erroneous transitions $(cccc)\to(cdcc)$ as the orange dashed arrow.
(c) Automaton representation of TFT-ATFT fully incorporating erroneous actions.
This representation is equivalent to the original history-based representation in (a).
}
\label{fig:tft_atft}
\end{figure}

Let us consider the iterated PD game between two players, say, Alice and Bob.
We assume that Alice has adopted TFT-ATFT, and its history-based representation
is shown in Fig.~\ref{fig:tft_atft}(a). The label of each node is the history
profile by Alice and Bob over the two previous rounds, denoted as
$A_{t-2}A_{t-1}B_{t-2}B_{t-1}$, where $A_t$ and $B_t$ mean Alice's and Bob's
actions at time $t$, respectively. This graph shows every possible transition
among the history profiles in the absence of the implementation error when
Alice is a TFT-ATFT player.

Alice normally behaves as a TFT player, and this behaviour is described by the
strongly connected component indicated by the green dashed rectangle in
Fig.~\ref{fig:tft_atft}(a). However, when she erroneously defects from mutual
cooperation, she switches her behaviour to ATFT. The history profile jumps from
$cccc$ to $cdcc$ by this error, and then Alice should defect once again as an
ATFT player. If both use TFT-ATFT, they quickly recover mutual cooperation
without being exposed to the risk of exploitation via the following sequence of
history profiles:
$cdcc \rightarrow ddcd \rightarrow dcdd \rightarrow ccdc \rightarrow cccc$.

The DFA minimization algorithm simplifies the graph to a great extent as shown
in Fig.~\ref{fig:tft_atft}(b). It has only four internal states which we have
labelled `4', `5', `0', and `1', respectively. It is the latter two, `0' and `1'
that describe the TFT behaviour, as we have already seen in
Fig.~\ref{fig:tft_as}.
When Alice erroneously defects from mutual cooperation, on the other hand, the
state jumps to `4', which belongs to the ATFT part, as indicated by the dashed arrow.
If both Alice and Bob
are TFT-ATFT players, they can safely recover the mutual cooperation at state
`0' via `5'.
The transition from `5' to `0' is crucial because Alice thereby accepts Bob's
punishment.

Although this automaton representation is meant to ignore error as we have mentioned, we depict a dashed arrow in Fig.~\ref{fig:tft_atft}(b) to indicate transition from $cccc \to cdcc$. This transition is the most important to understand how efficiency is satisfied because it is the only erroneous transition occurring with probability of $O(e)$ when two players adopt TFT-ATFT.
For the sake of completeness, Fig.~\ref{fig:tft_atft}(c) shows another automaton representation which fully takes into account erroneous actions.
State `0' and `1' in Fig.~\ref{fig:tft_atft}(b) are split into states `(11,0)' and `(1,7)' in Fig.~\ref{fig:tft_atft}(c), respectively.
We can fully reconstruct the original history-based representation of Fig.~\ref{fig:tft_atft}(a) from Fig.~\ref{fig:tft_atft}(c)
by checking every pair of successive arrows, noting that TFT-ATFT is a memory-two strategy.

\subsection*{Iterated three-person PG game}

\subsubsection*{Partially successful strategies}

\begin{figure}
\begin{center}
\includegraphics[width=0.9\textwidth]{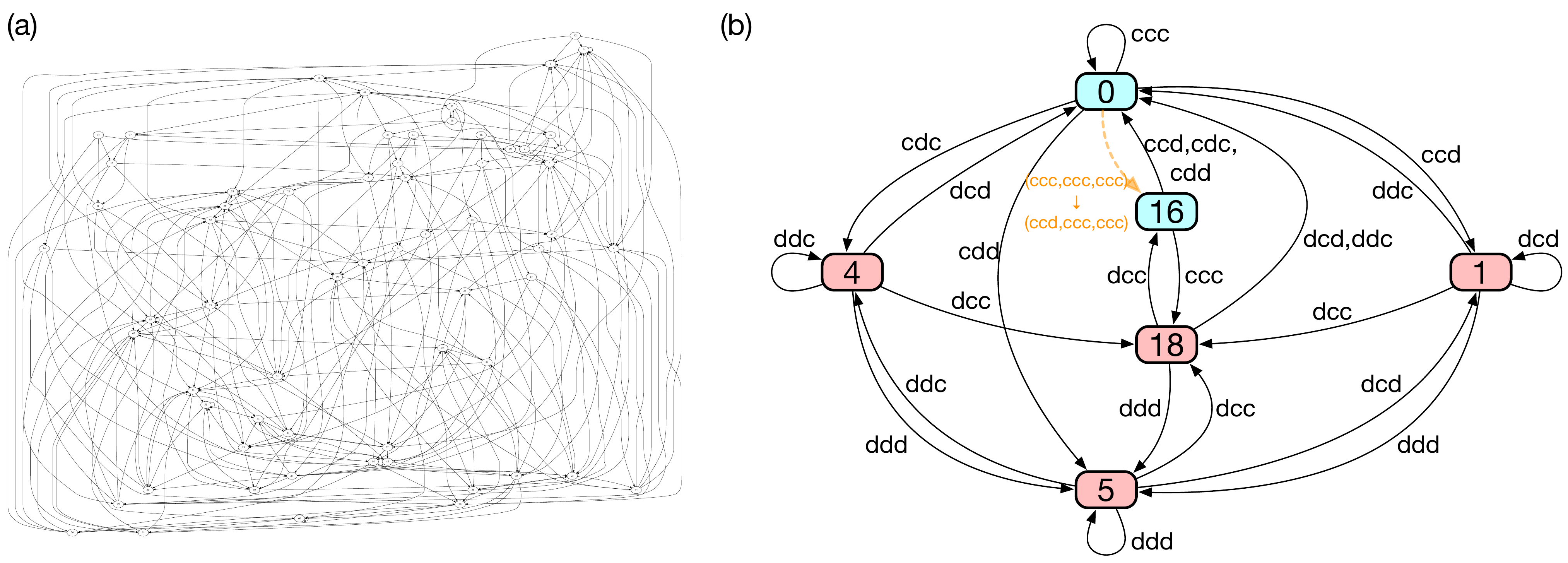}
\end{center}
\caption{
Two representations of one of the PS2's for the iterated three-person PG game.
(a) History-based representation.
As a memory-two strategy for a three-person game, it has $2^6 = 64$ nodes,
each of which has four outgoing links.
Note that it is practically impossible
to extract useful information from this representation even for $n=3$ and $m=2$.
(b) State-based representation.
The colour and the number of each node are depicted in the same way as in Fig.~\ref{fig:tft_atft}(b).
State `1' and `4' are symmetric under the exchange of the co-players.
Although error-caused transitions are not represented by this automaton, we
depict an erroneous action $(ccc,ccc,ccc)\to(ccd,ccc,ccc)$ for readers'
convenience as a dashed orange arrow.
}
\label{fig:ps2}
\end{figure}

Now, let us proceed to the iterated three-person PG game among Alice, Bob, and
Charlie. It has been proved for this game that
successful strategies are possible only when the memory length is
greater than two. However, it is instructive to begin with partially successful
strategies (PS2)~\cite{murase2018seven}, which are memory-two strategies with
defensibility, distinguishability, and partial efficiency. By partial
efficiency, we mean that the players achieve mutual cooperation with nonzero
probability $< 100\%$ in the limit of $e \to 0^+$. For example, TFT is partially
efficient.

By enumerating all the possible memory-two strategies, whose number is greater
than one trillion, we have discovered $256$ PS2's. Figure~\ref{fig:ps2} shows
one of them before and after the minimization. The history-based representation
needs $64$ nodes, which makes it difficult to interpret how the strategy works
by visual inspection [Fig.~\ref{fig:ps2}(a)]. On the other hand, its state-based
representation needs only $6$ nodes as demonstrated in Fig.~\ref{fig:ps2}(b).
Of course, some variations exist among PS2's, and the numbers of their internal
states are between $6$ and $8$ (or between $11$ to $15$ when erroneous actions are taken into account), but their overall structures are similar.

We can interpret the nodes in Fig.~\ref{fig:ps2}(b), representing the internal
states of this PS2, in the following way: Suppose that Alice is using this PS2.
The node labelled `0' means `full trust', and Alice can expect full cooperation
if this is her internal state. If one of her co-players, say Bob, defects from
full cooperation, Alice's state moves to `4', and her strategy prescribes
defection at this state. The meaning is obvious: She distrusts Bob. Cooperation
can nevertheless be recovered if Bob chooses $c$ whereas the other two players
punish him by $d$, whereby Alice's internal state becomes `full trust' again.
Another state labelled `1' can be interpreted in the same way, and the only
difference is that this time it is Charlie who defects from full cooperation.
Thus, states `1' and `4' are symmetric under the exchange of the co-players.
If both Bob and Charlie defect from full cooperation, Alice's internal state
changes to `5', which lies at the the bottom of Fig.~\ref{fig:ps2}(b). It
means that she is in despair because they are trapped in mutual defection.

So far, we have explained how we can interpret `0', `1', `4', and `5' in
Fig.~\ref{fig:ps2}(b). The interesting part is the other two states labelled `16'
and `18'. The former one, `16', corresponds to
$A_{t-2}A_{t-1}B_{t-2}B_{t-1}C_{t-2}C_{t-1} = cdcccc$, which is visited when
Alice defects erroneously from full cooperation, as indicated by the dashed arrow.
She has to choose $c$ at this
state, and she can go back to `0' by accepting defection from Bob or Charlie.
In plain words, therefore, we could say that Alice wants to make an apology at
this state. Similarly to TFT-ATFT, this apology plays an important role in
maintaining mutual cooperation in a noisy environment.

Alice can also visit `16' from `18' with a link of $dcc$. It is this state `18'
that makes it possible for Alice to provoke her co-players and test their
naivety: The loop between `16' and `18' implies that Alice can exploit Bob and
Charlie by alternating provocation ($d$) and apology ($c$) if they are
unconditional cooperators. This loop thus provides distinguishability for her
PS2.

\subsubsection*{Fully successful strategies}

\begin{figure}
\begin{center}
\includegraphics[width=0.95\textwidth]{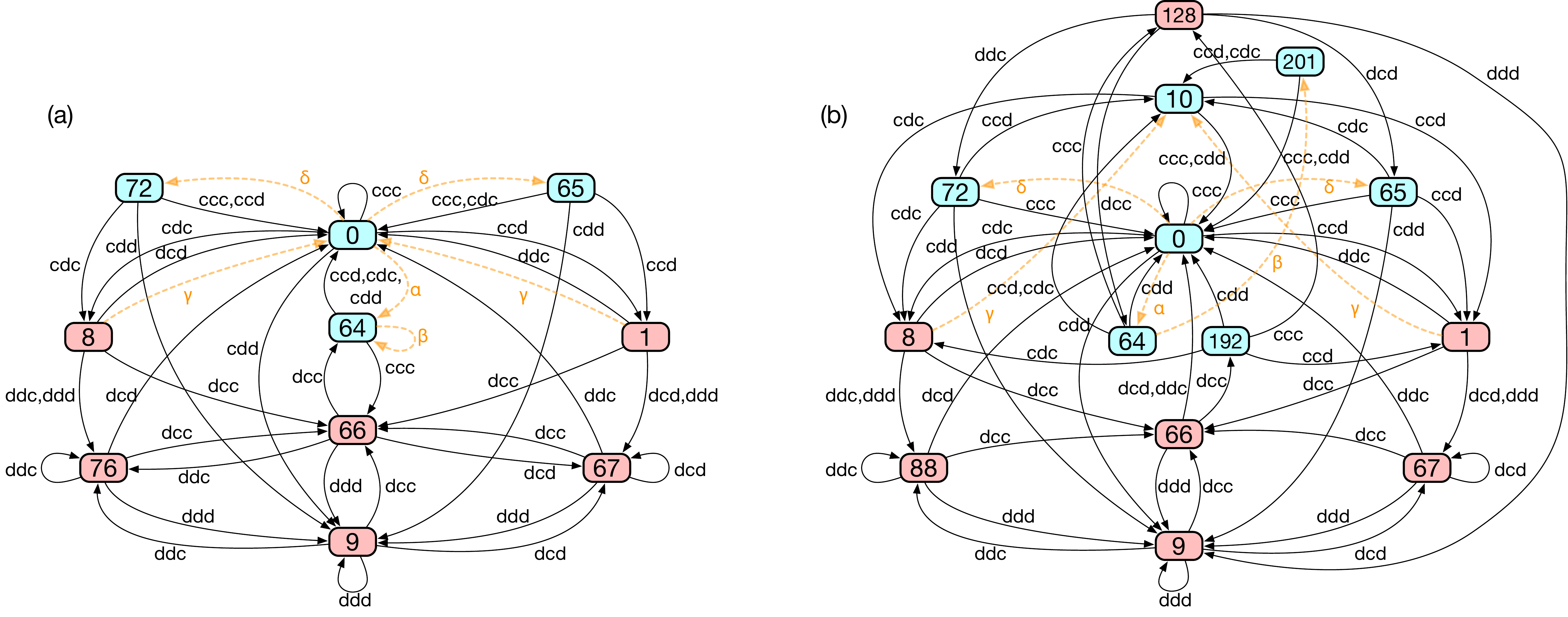}
\end{center}
\caption{
DFA minimization results of fully successful strategies for $n=3$ public-goods game.
One of the simplest and one of the most complex strategies are depicted in (a) and (b), respectively.
The labels and the colours are given in the same way as in Fig.~\ref{fig:tft_atft}(b).
The dashed orange arrows indicate erroneous actions occurring while recovering mutual cooperation from one- and two-bit errors.
The Greek letters correspond to the transitions shown in Fig.~\ref{fig:recovery_pattern}.
}
\label{fig:fs2}
\end{figure}

By modifying the $256$ PS2's, we have reported the same number of fully
successful strategies in the memory-three strategy
space~\cite{murase2018seven}. Their key difference from PS2's is that they
achieve full cooperation with probability $100\%$ in the limit of $e \to 0^+$.
To stress the difference, we will call them fully successful strategies
(FUSS's).
The DFA minimization process converts the FUSS's to automata with $10 \sim 14$
internal states (or $20 \sim 31$ states when errors are taken into account).
One of the simplest is depicted in Fig.~\ref{fig:fs2}(a).
Its similarity to Fig.~\ref{fig:ps2}(b) is striking, and we can immediately
recognize the states for full trust (`0'),  despair (`9'), apology (`64'), and
provocation (`66').
For reference, we also show the automaton representation for one of the most complex FUSS's in Fig.~\ref{fig:fs2}(b).
As shown in these automata representation, they show overall similar structures with sharing key mechanisms.
The same is true for the other FUSS's, and therefore, we focus on the simplest one in the following.

Obviously, some of its features are different from the above PS2:
First, this FUSS makes Alice more careful in distrusting one of her co-players.
Recall that we have interpreted state `4' as expressing Alice's distrustfulness
of Bob. It is now split into `8' and `76'. Due to this split, it takes one more
step to despair when one of the co-players defects. It means that the following
recovery path is possible even if Bob defects twice in a row:
\begin{equation}
0 \xrightarrow[]{cdc} 8 \xrightarrow[]{ddd} 76 \xrightarrow[]{dcd} 0,
\label{eq:bob2}
\end{equation}
whereas the same sequence of actions would only lead to Alice's distrustfulness
of Charlie for the PS2 shown in Fig.~\ref{fig:ps2}(b):
\begin{equation}
0 \xrightarrow[]{cdc} 4 \xrightarrow[]{ddd} 5 \xrightarrow[]{dcd} 1.
\end{equation}
The second difference is the appearance of `72' and `65', which have no
equivalents in Fig.~\ref{fig:ps2}(b). They are transient nodes with no incoming
links, which are reachable only by error. For example, node `72' represents a
history profile
\begin{equation}
A_{t-3}A_{t-2}A_{t-1}~B_{t-3}B_{t-2}B_{t-1}~C_{t-3}C_{t-2}C_{t-1} = ccd~ccd~ccc
\label{eq:simul}
\end{equation}
in binary, which means that Alice and Bob erroneously defected at
the previous round.
The states `65', `1', `67' are equivalent to `72', `8', `76', respectively,
when we swap the co-players Bob and Charlie.

\begin{figure}
\begin{center}
\includegraphics[width=0.9\textwidth]{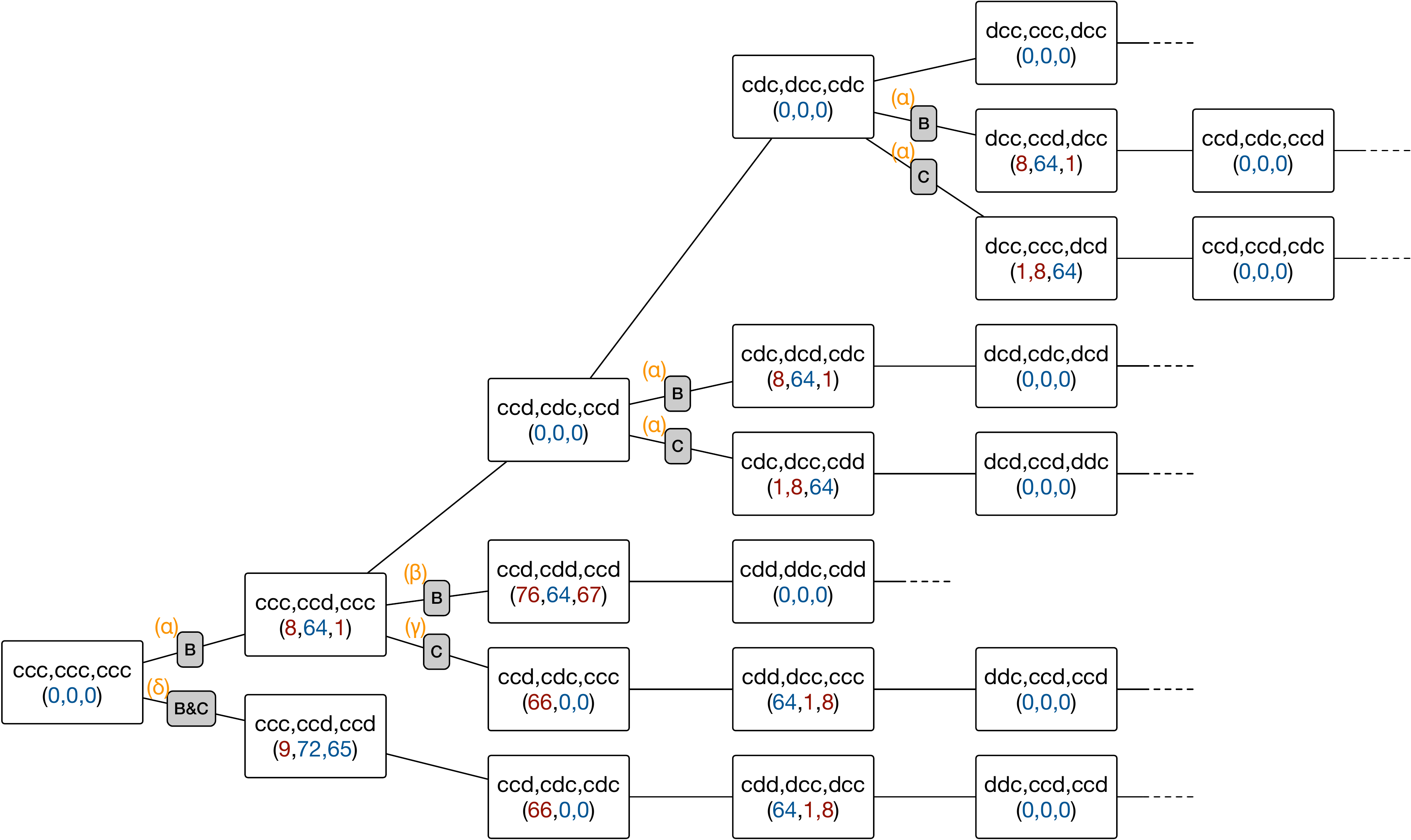}
\end{center}
\caption{
Paths for recovering mutual cooperation from one- and two-bit errors.
At each node, we have specified which history profile it represents, together
with the corresponding internal state in the state-based representation (see
the node labels in Fig.~\ref{fig:fs2}). The label of an internal state is
written in blue (red) if $c$ ($d$) is prescribed at the state. This figure
contains all the possibilities up to permutation of the players. The symbols `B'
and `C' in the shaded rectangles indicate who committed implementation error.
The orange Greek letters correspond to those in Fig.~\ref{fig:fs2}.
}
\label{fig:recovery_pattern}
\end{figure}

In fact, these additional four states are needed to make this strategy tolerant
against two-bit error, i.e., error of $O(e^2)$ either because it occurs one after another
or because it occurs to two players simultaneously.
Such tolerance is a necessary condition for full efficiency in
this three-person game~\cite{murase2018seven}. When Alice, Bob, and
Charlie have adopted this FUSS in common, we can show that the players recover
cooperation from every possible type of one- and two-bit error by enumerating
all the possible cases:
\begin{enumerate}
\item
Suppose that one of the FUSS players, say, Bob, committed an error. If $s_i$
denotes player $i$'s internal state (see Fig.~\ref{fig:fs2}), we
will have $(s_A, s_B, s_C) = (8,64,1)$, where Alice, Bob, and Charlie are
abbreviated to $A$, $B$, and $C$, respectively. It means that Bob will make an
apology (`64') by accepting punishment from Alice and Charlie. This recovers
cooperation as every player goes back to full trust (`0').
\item The FUSS can also correct two-bit error, which has three possibilities:
\begin{enumerate}
\item One player, say, Bob, commits error twice in a row.
\item Bob commits an error, and so does Charlie at the next round.
\item Two players, say, Bob and Charlie, commit error simultaneously.
\end{enumerate}
\end{enumerate}
Now, we will show that all these three types of two-bit error are corrected by
the FUSS:

(a) We have already seen from Eq.~\eqref{eq:bob2} that this FUSS
allows a recovery path along which $s_A$ changes as $0 \to 8 \to 76 \to 0$. The
question is whether the co-players' strategic interactions do not interrupt such
a path when they are using the same FUSS. In Fig.~\ref{fig:recovery_pattern}, we
keep track of the other players' states as well, according to which $(s_A, s_B,
s_C)$ changes as
\begin{equation}
(0,0,0) \xrightarrow[\text{Bob's error}]{cdc} (8,64,1) \xrightarrow[\text{Bob's
error}]{ddd} (76,64,67) \xrightarrow[]{dcd} (0,0,0)
\label{eq:2bit_a}
\end{equation}
for this type of two-bit error. The error is corrected.

(b) In the second case, Bob first defects in error from full cooperation.
Charlie is supposed to punish Bob by choosing $d$ at the next round, but he
mistakenly chooses $c$ instead. Up to this point, the players' internal states
have evolved as $(s_A, s_B, s_C) = (0,0,0) \to (8,64,1) \to (66,0,0)$.
Considering that state `66' has been interpreted as a decision to provoke the
co-players, we see that Charlie, after the mistaken $c$, appears to Alice as an
unconditional cooperator. After provoking Bob and Charlie by choosing $d$, Alice
wants to make an apology, and Bob and Charlie want to punish her provocation.
Their internal states thus correspond to $(64,1,8)$.
Understanding that Bob and Charlie are not
unconditional cooperators, Alice accepts their punishment, whereby everyone
returns to the full-trust state, i.e., $(s_A, s_B, s_C) = (0,0,0)$. The recovery
path is summarized as follows:
\begin{equation}
(0,0,0) \xrightarrow[\text{Bob's error}]{cdc} (8,64,1)
\xrightarrow[\text{Charlie's error}]{dcc} (66,0,0) \xrightarrow[]{dcc} (64,1,8)
\xrightarrow[]{cdd} (0,0,0).
\label{eq:case2b}
\end{equation}

(c) For the third case, we have already considered such simultaneous two-bit
error in Eq.~\eqref{eq:simul}. Alice falls into despair (`9') after Bob and
Charlie's simultaneous defection from full cooperation, and their states are
given as $(9,72,65)$. Bob and Charlie recognize their error and decide to
cooperate due to the existence of `72' and `65'. Alice tests them at $(66,0,0)$,
but we already know that this $(66,0,0)$ ends up with full trust [see
Eq.~\eqref{eq:case2b}]. The whole recovery path is thus given as follows:
\begin{equation}
(0,0,0) \xrightarrow[\text{Bob and Charlie's error}]{cdd} (9,72,65)
\xrightarrow[]{dcc} (66,0,0) \xrightarrow[]{dcc} (64,1,8)
\xrightarrow[]{cdd} (0,0,0).
\end{equation}
Although $c$ is prescribed by this FUSS at those newly added states, `72' and
`65', it does not violate the defensibility criterion because the states are
accessible only by its player's error, not by the co-players' intention.

To sum up, the FUSS in Fig.~\ref{fig:fs2} corrects every possible type of one-
and two-bit error and therefore exhibits full efficiency. All the recovery paths
discussed above are depicted together in Fig.~\ref{fig:recovery_pattern}.

\section*{Summary}\label{sec:summary}

In summary, we have investigated the working mechanism of successful strategies
for two- and three-person social dilemmas by converting them from history-based
representation to state-based one through the DFA minimization. The state-based
representation suggests how a player's internal state should interact with
observed actions to make the strategy successful.
The DFA minimization is especially effective when the number of players or the
memory length increases because the history profile expands exponentially with
them. It could also be useful when we evaluate the complexity of a strategy
according to the number of states in its automaton representation. We thus
believe that the method and the results presented in this paper serve as a solid
and indispensable step toward future explorations of successful strategies for
general $n$-person social dilemmas. Although successful strategies for $n > 3$
are yet to be found, they would share essential features or motifs
with the automata for the two- and three-person cases.

Such understanding of solving an iterated $n$-person social dilemma sheds
light on how to systematically manage collective action among self-interested
players by using the shadow of future. We may consider a number of social
phenomena such as voting and tax payments in the context of a social dilemma,
but the most prominent example is found in the fight against climate change: One
of the difficulties consists in the fact that it involves so many players who
individually favour defection from collective efforts to reduce greenhouse-gas
emissions. Even if a general consensus on cooperation exists on a global scale,
it might often be hard to deal with occasional defectors in practice because one
cannot tell if they have defected by accident or design. Our finding
nevertheless implies a possibility to devise a successful solution by sharpening
our intuition about when to retaliate against defection, when to accept
punishment by way of apology, and when to maintain cooperation.


\section*{Acknowledgements}
Y.M. acknowledges support from MEXT as ``Exploratory Challenges on Post-K
computer (Studies of multi-level spatiotemporal simulation of socioeconomic
phenomena)'' and from Japan Society for the Promotion of Science (JSPS) (JSPS
KAKENHI; Grant no. 18H03621). S.K.B. acknowledges support by Basic Science
Research Program through the National Research Foundation of Korea (NRF) funded
by the Ministry of Education (NRF-2020R1I1A2071670).
We acknowledge the hospitality at APCTP where part of
this work was done.

\section*{Author contributions statement}

S.K.B. conceived the work, and S.K.B. and Y.M. designed the research. Y.M.
created new software used in the work and analysed the results. All authors
reviewed and approved the manuscript.

\section*{Additional information}

The authors declare no competing interests.
The source code for this study is available at
\url{https://github.com/yohm/sim_automaton_successful_strategies}.

\end{document}